\documentstyle[12pt]{article}
\columnwidth\textwidth
\newcommand{\numu}{\nu_{\mu}}

\newcommand{\mmu}{m_{\mu}}
\newcommand{\me}{m_{e}}
\newcommand{\epsa}{\varepsilon_e}
\newcommand{\epsc}{\varepsilon_\nu}
\newcommand{\epsaa}{\varepsilon_e^2}
\newcommand{\epscc}{\varepsilon_\nu^2}
\newcommand{\glls}{g^S_{LL}}
\newcommand{\glrs}{g^S_{LR}}
\newcommand{\grls}{g^S_{RL}}
\newcommand{\grrs}{g^S_{RR}}
\newcommand{\gllsb}{g_{LL}^{S\star}}
\newcommand{\glrsb}{g_{LR}^{S\star}}
\newcommand{\grlsb}{g_{RL}^{S\star}}
\newcommand{\grrsb}{g_{RR}^{S\star}}
\newcommand{\gllv}{g^V_{LL}}
\newcommand{\glrv}{g^V_{LR}}
\newcommand{\grlv}{g^V_{RL}}
\newcommand{\grrv}{g^V_{RR}}

\newcommand{\glrvb}{g_{LR}^{V\star}}
\newcommand{\grlvb}{g_{RL}^{V\star}}

\newcommand{\glrt}{g^T_{LR}}
\newcommand{\grlt}{g^T_{RL}}
\newcommand{\glrtb}{g_{LR}^{T\star}}
\newcommand{\grltb}{g_{RL}^{T\star}}
\newcommand{\loga}{\log(1-x)}
\newcommand{\logb}{\log(\frac{m_{\mu}}{\mu})}
\newcommand{\logc}{\log(\frac{m_{\mu}}{m_e})}
\newcommand{\Li}{\mbox{Li}}
\begin{document}
\thispagestyle{empty}   
\noindent
ZU-TH 21/93\\
\\
ETHZ-ITP PR/93-05\\
\begin{center}
  \begin{Large}
  \begin{bf}
Effects of
non-standard couplings, radiative corrections
and neutrino masses on the
lepton spectra in $\mu$ and $\tau$ decays
\footnote{Work partially supported by Schweizerischer
Nationalfonds}
  \end{bf}
  \end{Large}
\end{center}
 \vspace{1.7cm}   
\begin{center}
  \begin{large}
    C. Greub and D. Wyler \\
    \vspace{0.5cm}
    Institut f\"ur Theoretische Physik,
    Universit\"at Z\"urich, Z\"urich, Switzerland \\
  \vspace{0.8cm}
    and  \\
  \vspace{0.8cm}
    W. Fetscher            \\
    \vspace{0.5cm}
    Institut f\"ur Teilchenphysik (ITP), ETH
Z\"urich, \\
    CH-5232 Villigen, Switzerland       \\
  \end{large}
\vspace{2.0cm}
\end{center}
\begin{abstract}
We investigate the combined effect of neutrino masses
and
non-standard couplings on the various lepton spectra in
$\mu$ and  $\tau$ decays. We
emphasize
the energy spectra of the neutrinos, which
can be
measured by secondary reactions in the new KARMEN
experiment
and hence will yield novel information on deviations
from the $V-A$
structure. To constrain these couplings, QED radiative
corrections have to be taken into account.
We evaluated them and found small corrections to the
neutrino spectra.
\end{abstract}
\newpage
\section{Introduction}
The decay $\mu^+ \to e^+ \bar{\nu}_\mu \nu_e$ is  of
special interest
to both theory and experiment. It is purely leptonic and
therefore
one can investigate the properties of the weak
interactions
without QCD complications. In the standard model,
the $V-A$ form
of the charged current follows   {\it by
construction}.
Recently, Fetscher, Gerber and Johnson (FGJ)
\cite{FJG} showed that $V-A$ in fact emerges
with remarkable
accuracy
from a small set of experiments;
nevertheless, some of the non- $V-A$ couplings could
still be
substantial. In all of the muon decay experiments up to
then,
only muon and electron properties had been measured and
an
ambiguity
in nailing down the correct coupling had remained. FGJ
had found
that
it can be resolved by using the cross section of the
inverse
reaction $\numu e^- \to \mu^- \nu_e$ \cite{FJG}.

A recent experiment \cite{Bod} allows for the first time
to include the information about the $\nu_e$ from muon
decay, in which the
neutrinos from  $\mu$ decay undergo a secondary
interaction,
either $^{12}C(\nu_e,e^-)^{12}N(g.s.)$ by a charged
current process or
$^{12}C(\nu,\nu)^{12}C$ by a neutral current scattering
($\nu=\nu_e,\bar{\nu}_\mu$). This allows to measure the
neutrino spectra, both
for the electron neutrino (charged and neutral currents)
and the muon
neutrino (neutral current). As was argued by Fetscher
\cite{F,Fe93}, these spectra
are sensitive to new interactions which may cause slight
deviations from the
$V - A$ behaviour and thus offer an alternative to
scattering data.
In order to make these meaningful, QED radiative
corrections, which are
expected to be of the same order of magnitude as
possible effects from new
physics, must be determined. We therefore compute the
radiative corrections to
the $\nu_e$ spectrum.

Also, since $m_{\nu_\mu} < 270 keV (90\% c.\ell.)$
\cite{Abe}, its
effects could be comparable to those of $m_e$. Even
larger are the possible
effects of $m_\mu$ and of $m_{\nu_\tau} <
31 MeV (90\% c.\ell.)$
\cite{Alb} in $\tau$ decays. We have therefore also
evaluated the effects of
these masses in spectra and rates; we also mention
briefly their influence on
polarization measurements.
Purely leptonic decays with massive neutrinos and arbitrary
Lorentz structure were considered by Shrock \cite{shrock}
on a very general basis. Here we focus on those effects of
non-zero masses of the $\nu_\mu$ and $\nu_\tau$ on
spectra and rates which could become measurable in the near
future. These effects are due to interference terms and are
linear in the neutrino masses. We express the corresponding
decay parameters in terms of chiral coupling constants given
below, which are best suited to describe the interference.

In order to fix our notation, we write down the general
form of the
Hamiltonian for the $\mu$-decay \cite{FJG,FG}
\footnote{ This review contains all necessary background
information on
  $\mu$ decay}
\begin{equation}
\label{matel}
{\cal H} = \frac{4 G_F}{\sqrt{2}} \,
\sum_{\gamma=S,V,T;\varepsilon,\delta=R,L}
\, \left\{ g^\gamma_{\varepsilon \delta} \,
\left[ \bar{e}_\varepsilon
\, \Gamma^\gamma \, (\nu_e)_n \right]  \, \left[
(\bar{\nu}_\mu)_m
\, \Gamma_\gamma \, \mu_\delta \right] + h.c. \right\}
\quad .
\end{equation}
Here, $\gamma$ labels the type of interaction (scalar,
vector, tensor) and
$\varepsilon$ and $\delta$ the chirality of the charged
leptons. Those of the
neutrinos, $n$ and $m$ are uniquely fixed by $\gamma$,
$\varepsilon$ and
$\delta$. The couplings $g^\gamma_{\varepsilon \delta}$
are
normalized as in eq.
(2.61) of ref. \cite{FG} and their present values or
bounds
are also given in
\cite{FG}. In this notation, $g^V_{LL}=1$, all other
$g^\gamma_{\varepsilon
\delta}=0$, corresponds to the standard model. The
non-standard couplings
$g^\gamma_{\varepsilon \delta}$ arise in extensions of
the standard model
such as left-right models, extended Higgs structure,
dileptons etc. These
effects, calculated for $\mu$ decay, can easily be
extended to leptonic
$\tau$-decays by simply substituting $\mu \to \tau$ and
$e \to e \, \mbox {or} \,
\mu$.
\section{Mass effects on the decay rate and on the $e^+$
energy
spectrum}
It is straightforward to calculate the rate for
$\mu$ decay including all mass effects.
In the following, $m_\mu$, $m_e$ and $m_\nu$ denote
the masses of the
$\mu^+$, $e^+$ and $\bar{\nu}_\mu$; the mass of the
$\nu_e$
is always neglected in this paper.
Omitting also
terms of order $\alpha \cdot (m_e/m_\mu)$,
$\alpha \cdot (m_\nu/m_\mu)$ and $\alpha \, \cdot
g^\gamma_{\varepsilon \delta}$
(except $g^V_{LL}$ of course), we obtain
\begin{equation}
\label{gamtot}
\Gamma_\mu = \frac{G_F^2 \, m_\mu^5}{192 \pi^3} \,
\left( 1
- 8  \epsaa - 8  \epscc  + 4 \eta \epsa + 4 \lambda \,
\epsc + 8
\sigma \, \epsa \epsc
\right) \, f_W f_r  \quad ,
\end{equation}
where $\epsa = m_e/m_\mu$, $\epsc = m_\nu/m_\mu$ and
$f_W$ and $f_r$ are weak and electromagnetic corrections
\begin{eqnarray}
\label{weakcorr}
f_W &=& 1 + \frac{3}{5} \, \left( \frac{m_\mu}{m_W}
\right)^2
\nonumber \\
f_r &=& 1 - \frac{\alpha}{2 \pi} \, \left( \pi^2 -
\frac{25}{4}
\right) \qquad .
\end{eqnarray}
In eq. (\ref{gamtot}), additional terms of order
$O(\varepsilon^3)$ and $O(\varepsilon^3
\log(\varepsilon))$ have
been neglected.
The quantities $\eta$, $\lambda$ and $\sigma$ are
functions of the
coupling constants $g^\gamma_{\varepsilon \delta}$:
\begin{equation}
\label{eta}
\eta = \frac{1}{2} Re \left\{ \gllv \grrsb + \grrv
\gllsb +
\glrv \, ( \grlsb + 6 \grltb ) +
\grlv \, ( \glrsb + 6 \glrtb ) \right\}
\end{equation}
\begin{equation}
\label{lambda}
\lambda = \frac{1}{2} Re \left\{ \glls \glrsb + \grrs
\grlsb-
2 \, \grrv \grlvb - 2 \,  \gllv \glrvb \right\}
\end{equation}
\begin{equation}
\label{sigma}
\sigma = \frac{1}{2} Re \left\{  \grrs \glrvb + \glls
\grlvb +
\gllv \, ( \grlsb - 6 \grltb ) +
\grrv \, ( \glrsb - 6 \glrtb ) \right\} \ .
\end{equation}

In the standard model $\eta=\lambda=\sigma=0$. With the
experimental value in
$\gllv \approx 1$ we can neglect terms of second order
in
$g^\gamma_{\varepsilon
\delta}$  ($g^\gamma_{\varepsilon\delta} \ne \gllv$) and
obtain
\begin{equation}
\label{etaapp}
\eta = \frac{1}{2} Re \left\{ \grrs \right\}
\end{equation}
\begin{equation}
\label{lambdaapp}
\lambda = - Re \left\{ \glrv \right\}
\end{equation}
\begin{equation}
\label{sigmaapp}
\sigma = \frac{1}{2} Re \left\{
 \grls - 6 \grlt \right\} \ .
\end{equation}

We note that the present value of $G_F$ has been derived
from the muon decay
width $\Gamma_\mu$ (resp. from the muon lifetime
$\tau_\mu$ ) assuming a
pure $V-A$ interaction. Without this assumption the
experimental error
on $G_F$ increases by a factor of 20 due to the present
experimental error
on $\eta$~\cite{FG}. The $\nu_\mu$ interference term
$\lambda\varepsilon_\nu$
can lead  to comparable errors on $G_F$. Both terms
play an even larger role
when deriving $G_F$ from leptonic $\tau$ decays to test
the universality of
the charged weak interaction.

Similarly we calculate the $e^+$ energy spectrum
including the effects of the
$e^+$ and $\bar{\nu}_\mu$ masses. The radiative
corrections to this spectrum
were explicitely given in ref. \cite{KS} for the $V-A$
case in the relevant
limit $m_e,m_\nu \to 0$ and we therefore do not discuss
them.

Introducing the reduced $e^+$ energy $x_e=E_e/W_e$ (with
$W_e \doteq
\frac{m_\mu}{2} \, (1 + \epsaa -\epscc)$) which varies
in the interval
\begin{equation}
\label{ekinematics}
x_e^0 \le x_e \le 1 \quad , \quad \mbox{with} \quad
x_e^0=
(2 \epsa)/(1+ \epsaa - \epscc) \quad ,
\end{equation}
the spectrum $d\Gamma/dx_e$ reads
\begin{equation}
\label{espectrum}
\frac{d \Gamma}{dx_e} = \frac{G_F^2 \, W_e^4 \,
m_\mu}{\pi^3} \,
\sqrt{(x_e)^2 - (x_e^0)^2} \,
\left[ H_1 + \frac{2}{9} \rho H_2 + \eta
\epsa  H_3 +
\lambda \epsc  H_4
+ \sigma  \epsa  \epsc  H_5  \, \right] \quad .
\end{equation}
With $t = (1 +
\epsaa)(1-x_e) + x_e \epscc $ the functions $H_1, ...,
H_5$
are given by
\begin{eqnarray}
\label{hfunctions}
H_1 &=&  \frac{(1 - t + \epsaa) \, (1 - x_e)^2}{t}
\nonumber \\
H_2 &=& \left\{ 4 t^3 -t^2 ( 5 + 5 \epsaa + \epscc) + t
\left[
(1 - \epsaa)^2 - \epscc ( 1 + \epsaa) \right] + \right.
\nonumber \\
&& \hspace{1cm} \left. + 2 \epscc (1- \epsaa)^2 \,
\right\}
\, \frac{(1-x_e)^2}{t^3} \nonumber \\
H_3 &=& \frac{2 (1-x_e)^2 }{t} \nonumber \\
H_4 &=& \frac{(1-t-\epsaa) \, (1-x_e)^2 }{t^2}
\nonumber \\
H_5 &=& \frac{(1+t-\epsaa) \, (1-x_e)^2 }{t^2} \quad .
\nonumber \\
\end{eqnarray}
The parameters
$\eta$, $\lambda$ and $\sigma$ are listed in eqs.
(\ref{eta}),
(\ref{lambda}) and (\ref{sigma}), respectively;
the Michel parameter $\rho$ is
\begin{eqnarray}
\label{rho}
\rho &=& \frac{3}{16} \, \left\{
|\glls|^2 + |\glrs - 2 \glrt|^2 + |\grls - 2 \grlt|^2
+
\right . \nonumber \\
&& \hspace{1cm} \left.
+ |\grrs|^2 + 4 |\gllv|^2 + 4 |\grrv|^2 \right\}
\quad .
\end{eqnarray}
\section{The $\nu_e$ energy spectrum}
With the new generation of experiments (KARMEN)
\cite{Bod} the
electron neutrino energy
spectrum can be measured with a precision of the order
of a few percent. This
will yield improved limits to non-standard contributions
to muon decay.
While the tree level form was given by Fetscher
\cite{F,Fe93}, a sensible test of
the new interactions requires inclusion of the radiative
QED corrections to
the standard contribution and of the mass effects of the
$e^+$ and the
$\bar{\nu}_\mu$.
\subsection{Radiative QED corrections}
The radiative corrections to order $\alpha$ consist of
two parts:
Bremsstrahlung processes where a real photon is radiated
off the electron or
the muon and the virtual corrections (self-energies of
the electron and the
muon, and the $\mu - e$ vertex correction) whose
interference with the zeroth
order process also yields a contribution of order
$\alpha$. Since the
bremsstrahlung photon from the $\mu$ decay is not
detected in the experimental
setup, we integrate over all its frequencies.

These corrections have been considered long ago by
Behrends, Finkelstein,
Kinoshita and Sirlin \cite{KS,BFS}, who applied them to
the positron energy
distribution. The remarkable fact is, that although the
four-Fermi interaction
is non-renormalizable, all ultraviolet divergences can
be absorbed in the
mass of the charged leptons for $V,A$ interactions.

As we look for  a spectrum which is not explicitly given
in the literature,
we did not make use of the previous work, but performed
a new calculation.
While taking into account the effects of $m_e$ and
$m_\nu$ at tree level,
we neglect these masses in the radiative corrections.

Both the ultraviolet and the infrared singularities are
regularized
dimensionally. While the limit $m_\nu \to 0$ can be
taken at the very
beginning of the calculation, we note that individually
both, the
bremsstrahlung and the virtual corrections are plagued
with mass singularities
for $m_e =0$.
We therefore use a finite electron mass as a regulator.
Only when adding
virtual and bremsstrahlung corrections the limit $m_e
\to 0$ is finite
due to the KLN theorem \cite{KLN}. As one of us has
discussed the details of
the regularization procedure in a similar process at
length in ref.
\cite{HPA}, we only give the results, separating into
bremsstrahlung and
virtual corrections to the electron neutrino spectrum.

We define the dimensionless electron neutrino energy $x$
by
\begin{equation}
\label{range}
E_{\nu_e} = m_{\mu} \cdot \frac{x}{2} \quad , \quad x
\in [0,1]
\footnote{Note, that we neglect $m_e$ and $m_\nu$ in
radiative
corrections in contrast to section 3.2, where tree-level
processes
are discussed} \ .
\end{equation}
The virtual correction is split into an infrared
singular and an infrared
finite part. We work in $d=4-2\varepsilon$ dimensions;
the
singular part is by
definition the term proportional to $1/\varepsilon$.
Writing
\begin{equation}
\label{virtsplit}
\frac{d\Gamma^{virt}}{dx} =
\frac{d\Gamma^{virt}_{sing}}{dx}
+ \frac{d\Gamma^{virt}_{fin}}{dx} \quad ,
\end{equation}
we have
\begin{eqnarray}
\label{virtsing}
\frac{d\Gamma^{virt}_{sing}}{dx} &=& - \frac{G_F^2 \,
\mmu^5 \,
\alpha}{16 \, \pi^4} \, \frac{\Gamma(1-\varepsilon) \,
(4
\pi)^{3
\varepsilon}}{\Gamma^2(2-2\varepsilon)} \, x \, (1-x)
\times
\nonumber
\\
&& \times \left[ (1-x) \, \loga + 2 x - x \logc \right]
\,
\frac{1}{\varepsilon}
\end{eqnarray}
\begin{eqnarray}
\label{virtfin}
\frac{d\Gamma^{virt}_{fin}}{dx} &=&  \frac{G_F^2 \,
\mmu^5
\,
\alpha}{32 \, \pi^4} \,
(1-x) \times \nonumber \\
&& \times \left[5 x (1-x) \, \log^2(1-x) + (1 - 6 x + 9
x^2)
\loga + \right. \nonumber \\
&& + 4x(1-x) \loga \log(x) + 8 x^2 \log(x) + \nonumber
\\
&& \left.+4x(1-x) \Li(x) + x -13 x^2 + A(x) \right]
\qquad .
\end{eqnarray}
Here the Spence function $\Li(x)$ is defined by
\[ \Li(x) = - \int_{0}^{x} \, \frac{dt}{t} \, \log(1-t)
\quad . \]
The term $A(x)$ in eq. (\ref{virtfin}) contains the mass
singularities $(m_e \to 0)$ and the terms involving the
renormalization scale $\mu$ introduced in dimensional
regularization:
\begin{eqnarray}
\label{a}
A(x) &=& 12 x (1-x) \, \loga \, \logb + 2 x^2
\log^2(\frac{\mmu}{\me}) \nonumber \\
&& -2 x^2 \, \loga \, \logc -12 x^2 \, \logb \, \logc
\nonumber
\\
&& - 4 x^2 \, \logc \, \log(x) + x^2 \, \logc + 24 x^2
\, \logb
\qquad .
\end{eqnarray}
Similarly, the bremsstrahlung corrections are split into
an infrared singular
and an infrared finite part.
\begin{equation}
\label{bremssplit}
\frac{d\Gamma^{brems}}{dx} =
\frac{d\Gamma^{brems}_{sing}}{dx}
+ \frac{d\Gamma^{brems}_{fin}}{dx} \quad ,
\end{equation}
with
\begin{equation}
\frac{d\Gamma^{brems}_{sing}}{dx} =
- \frac{d\Gamma^{virt}_{sing}}{dx} \quad \mbox{and}
\end{equation}
\begin{eqnarray}
\label{bremsfin}
\frac{d\Gamma^{brems}_{fin}}{dx} &=& - \frac{G_F^2 \,
\mmu^5
\,
\alpha}{192 \, \pi^4} \,
(1-x) \, \left[
6 \, (5x-4x^2) \, \log^2(1-x) + \right. \nonumber \\
&& + (11 -28x + 62x^2) \, \loga + 24 x (1-x) \, \loga \,
\log(x)
\nonumber \\
&&+ 48 x^2 \, \log(x) + 12 x (2-x) \, \Li(x) + 24 x^2
\Li(1) +
\nonumber \\
&& \left.+11x -\frac{175}{2} x^2 + 6 A(x) \, \right]
\quad .
\end{eqnarray}
Adding the bremsstrahlung and the virtual corrections we
get the total
$O(\alpha)$ correction to the electron neutrino energy
spectrum, which in the
limit $m_e \to 0$ can be written as
\begin{equation}
\label{final1}
\frac{d\Gamma^{corr}}{dx} = - \frac{d\Gamma^0}{dx}
\,  \frac{\alpha}{2
\pi} \, G(x)  \quad , \quad
\frac{d\Gamma^0}{dx}  =
\frac{G_F^2 \, \mmu^5
\, x^2 \, (1-x)}{16 \, \pi^3} \quad .
\end{equation}
Here $(d \Gamma^0/dx)$ is the standard model spectrum
without radiative
corrections in the limit $m_e=m_\nu=0$. The function
$G(x)$ reads
\begin{eqnarray}
\label{final2}
G(x) &=& \frac{1}{12} \, \left[ 12 \log^2(1-x) + \left(
\frac{10}{x^2} + \frac{16}{x} + 16 \right) \, \log(1-x)
+ \right.
\nonumber \\
&& \left. + 48 \Li(1) + 24 \Li(x) + \frac{10}{x} - 19
\right]
\qquad .
\end{eqnarray}
As expected, $(d \Gamma^{corr}/dx)$ is free from mass
singularities.
It is very small and in fact the corrections are
invisible
in a plot of $(d\Gamma/dx)/\Gamma$.
In contrast, the corrections to the $e^+$
energy spectrum, given in
Fig. 1 of ref. \cite{KS}, are much bigger. The reason is
easy to understand.
While $(d \Gamma^{corr}/dx)$ in eq. (\ref{final1}) is
small, the corrections
to the $e^+$ energy spectrum contain rather large
logarithms of the form
$\log^2(m_e/m_{\mu})$ and $\log(m_e/m_{\mu})$.

The result (\ref{final2}) is of course also applicable
to QCD corrections to
the $e^+$ distribution in semileptonic charm decay in
the limit $m_s \to 0$.
This has been  considered in ref. \cite{Cabibbo};
however, their function $G$
does not agree with eq. (\ref{final2}). On the other
hand we have checked numerically
and analytically that our results agree with those of
ref.
\cite{ali_pietarinen} if $m_s$ vanishes.
\subsection{Mass effects on the $\nu_e$ energy spectrum}
As the secondary reaction is hardly sensitive to a
right-handed
$\nu_e$, we only calculate the production of a
left-handed
$\nu_e$.
Using the scaled energy $x$ of
the $\nu_e$ defined in eq. (\ref{range}),
we get
\begin{equation}
\label{nuespectrum}
\frac{d \Gamma_L}{dx} = \frac{G_F^2 \, m_\mu^5}{16 \,
\pi^3} \,
Q_L^{\nu_e} \,
\left\{ G_1 + \omega_L  G_2 + \eta_L \epsa
 G_3 + \lambda_L  \epsc G_4
+
\sigma_L  \epsa \epsc G_5 \right\}   \quad .
\end{equation}
$x$ now varies in the interval $ 0 \le x  \le 1 - (\epsa
+ \epsc)^2$.
The functions $G_1, ..., G_5$ are most
easily expressed in the variables
$r=(1-x)$, $\delta = (\epsaa - \epscc)$, $\kappa =
(\epsaa + \epscc)$ and
$\xi = \sqrt{r^2 - 2 \kappa r + \delta^2}$:
\begin{eqnarray}
\label{gfunctions}
G_1 &=& \frac{(1-r)^2 \, (r - \kappa) \, \xi}{r}
\nonumber \\
G_2 &=& \frac{2}{9} \, \frac{(1-r)^2 \, \xi}{r^3} \,
\left[ -4r^3 + r^2 \left( 1 + 5 \kappa \right) + r
\left( \kappa - \delta^2 \right) - 2 \delta^2 \right]
\nonumber \\
G_3 &=& \frac{(1-r)^2 \, \xi}{r^2} \, (r - \delta)
\nonumber \\
G_4 &=& \frac{(1-r)^2 \, \xi}{r^2} \, (r + \delta)
\nonumber \\
G_5 &=& \frac{2 \, (1-r)^2 \, \xi}{r}
\quad .
\end{eqnarray}
The parameters $Q_L^{\nu_e}$, $\omega_L$, $\eta_L$
\cite{Fe93} and the
 new terms $\lambda_L$ and $\sigma_L$ for left-handed
$\nu_e$
read
\begin{eqnarray}
\label{couplingleft}
Q_L^{\nu_e} &=& \frac{1}{4} \, \left\{
|\grls|^2 + |\grrs|^2 + 4 \, |\gllv|^2 + 4 \, |\glrv|^2
+
12 \, |\grlt|^2 \right\}
\nonumber \\
\omega_L &=& \frac{3}{16} \, \left\{
|\grrs|^2 + 4 \, |\glrv|^2 + |\grls + 2 \, \grlt|^2
\right\}/Q_L^{\nu_e} \nonumber \\
\eta_L &=& \frac{1}{2} \, Re \left\{
\gllv \grrsb + \glrv \,(\grlsb + 6 \grltb )
\right\}/Q_L^{\nu_e} \nonumber \\
\lambda_L &=& \frac{1}{2} \, Re \left\{
\grrs \grlsb - 2 \, \gllv \glrvb
\right\}/Q_L^{\nu_e} \nonumber \\
\sigma_L &=& \frac{1}{2} \, Re \left\{
\glrv \grrsb + \gllv \,(\grlsb - 6 \grltb )
\right\}/Q_L^{\nu_e} \quad .
\end{eqnarray}
The radiatively corrected spectrum $d\Gamma_L^{tot}/dx$
is now obtained
by combining eqs. (\ref{nuespectrum}) and (\ref{final1})
\begin{equation}
\label{alles}
\frac{d\Gamma_L^{tot}}{dx} = \frac{d\Gamma_L}{dx} +
\frac{d\Gamma^{corr}}{dx} \quad .
\end{equation}
There is an apparent problem in connection with the
range of the variable $x$
in eq. (\ref{alles}). In the radiative corrections
$d\Gamma^{corr}/dx$, which
have been calculated for $m_e=m_\nu=0$, $x$ varies
between 0 and 1, whereas in
the tree level contribution $d\Gamma_L/dx$ $x$ varies
between 0 and
$(1-(\epsa+\epsc)^2)$ due to the finite masses $m_e$ and
$m_\nu$. However, to
the precision we are working, we are allowed to restrict
$x$ to the common
range $0 \le x \le (1-(\epsa+\epsc)^2)$ in eq.
(\ref{alles}).

In Fig. 1a  we compare the radiative corrections (eq.
(\ref{final1})) to the effect of a non-zero $\omega_L$,
which is obtained from eq.
(\ref{nuespectrum}) by retaining
only the term proportional to $\omega_L$
in the bracket.
For the present experimental upper limit $\omega_L =
0.1$
\cite{Dre}
(and choosing $Q_L^{\nu_e} = 1$),
we see that these are much larger.
Only for $\omega_L \simeq 0.01$ radiative corrections
become comparable as illustrated in Fig. 1b.
\section{The $\bar{\nu}_\mu$ energy spectrum}
We also give the energy spectrum of the muon
antineutrino
which can be measured
through neutral current reactions at the secondary
vertex. While one expects
the $(V - A)$ term to yield the same tree level spectrum
(in the equal mass
limit for $e^+$ and $\bar{\nu}_\mu$) as the one of the
positron, the other
terms may distort it in different ways. Keeping the
$\bar{\nu}_{\mu}$-mass
(bounded only by about $m_e/2$ \cite{Abe}), one obtains
for the energy
distribution of the $\bar{\nu}_\mu$
\begin{equation}
\label{numuspectrum}
\frac{d \Gamma}{dy} = \frac{G_F^2 \, m_\mu^5}{16 \,
\pi^3} \,
\left\{ K_1 + \tilde{\omega} K_2 +
\eta \epsa K_3 + \lambda  \epsc
K_4 +
\sigma  \epsa  \epsc  K_5
\right\} \quad ,
\end{equation}
where the scaled muon antineutrino energy $y$
\[ y = \frac{2 E_{\nu_\mu}}{m_\mu \, (1 + \epscc)} \]
varies in the interval
\[ \frac{2 \epsc}{1 + \epscc} \le y \le
\frac{1 + \epscc - \epsaa}{1 + \epscc}  \quad . \]
The functions $K_1, ..., K_5$ are easily written in the
variables $s=(1-y) \,
(1 + \epscc)$,
$\zeta = \sqrt{s^2 - 2 (1 + \epscc) s + (1 - \epscc)^2}$
as follows:
\begin{eqnarray}
\label{kfunctions}
K_1 &=& \frac{\zeta}{s} \, (s - \epsaa)^2 \,
(1 - s + \epscc) \, (1 + \epscc)
\nonumber \\
K_2 &=& \frac{2 \zeta}{9 s^3} \,
(s - \epsaa)^2 \, (1 + \epscc) \times
\nonumber \\
&&   \times
\left[ 4s^3 - s^2 \left( 5 + 5 \epscc + \epsaa \right) +
s
\left( (1 - \epscc)^2 - \epsaa (1 + \epscc)
\right)
+ 2 \epsaa \, (1 - \epscc)^2
\right]
\nonumber \\
K_3 &=& \frac{\zeta}{s^2} \, (1- s - \epscc) \,
(s - \epsaa)^2 \, (1 + \epscc)
\nonumber \\
K_4 &=& \frac{2 \zeta}{s} \, (s - \epsaa)^2 \,
(1 + \epscc)
\nonumber \\
K_5 &=& \frac{\zeta }{s^2} \,
(1 + s - \epscc) \, (s - \epsaa)^2 \, (1 + \epscc)
\quad.
\end{eqnarray}
The couplings $\eta$,
$\lambda$ and $\sigma$ are given in eqs. (\ref{eta}),
(\ref{lambda}) and (\ref{sigma}), respectively, while
we get for $\tilde{\omega}$
\begin{equation}
\label{omegatilde}
\tilde{\omega} = \frac{3}{4} \, \left\{
|\gllv|^2 + |\glrv|^2 + |\grlv|^2 + |\grrv|^2 +
4 \, |\glrt|^2 + 4 \, |\grlt|^2 \right\} \quad .
\end{equation}
Note that $\tilde{\omega}$ is the analogon of the usual
$\rho$-parameter.
\section{Remarks}
We have calculated the effects of electromagnetic
corrections,
masses and non-standard couplings
on the rate and spectra in $\mu$ and  $\tau$ decays.
Due to the absence of large logarithms, the radiative
corrections
to the neutrino energy spectra are very small, unlike
those
to the spectra of charged leptons.
Therefore, the neutrino distributions are in fact
suitable
for bounding new interactions (see Figs. 1a,b).
The new experiments (KARMEN) \cite{Bod} will
measure these spectra in $\mu$ decay
with high precision;
preliminary results \cite{Dre}
are indeed very encouraging.
Like for the electron, neutrino mass effects are very
small
in decay rates and spectra
(the terms proportional to $\lambda$ are  of relative
order
$O(10^{-4})$ in $\mu$ decay).
But as is known \cite{FJG} for the electron
mass term proportional to $\eta$, polarization
measurements could
drastically enhance the sensitivity.  For
$\tau$-decay, however, the mass effects are larger, as
$(\eta \varepsilon_e)$ is replaced by
$(\eta^\tau m_\mu/m_\tau)$, where $\eta^\tau$ is defined
for
$\tau$-decays as in eq. (\ref{eta}).  We note that
bounds on $\eta$, $\lambda$ etc. at the level of
$O(\alpha^2)$
are  softened by the yet
uncomputed electromagnetic two loop  corrections to
the standard model contribution
and the one loop corrections
to the non-standard contributions. With the presently
attainable experimental
precision, however, measurements in muon and leptonic
tau decays
continue to
yield important contributions to our understanding
of
the fundamental interactions.

\section*{Figure captions}
\subsection*{Figure 1}

Contributions to the $\nu_e$ spectrum $d\Gamma/dx$ from
muon decay due to
non-standard interactions (Fig. 1a: $\omega_L=0.1$, Fig.
1b: $\omega_L=0.01$)
and to radiative corrections as a function of the
reduced neutrino energy
$x=E_\nu/m_\mu$. Radiative corrections have to be taken
into account when
evaluating experimental spectra at the $\%$-level, but
they play a minor role
in the region $x>0.9$ where the sensitivity for
$\omega_L$ is greatest.
\end{document}